\documentclass{cimento}
\usepackage[british]{babel}

\title{Accurate predictions for top-quark pair production at the LHC}
\author{
  S.~Ferrario Ravasio
  \thanks{Presented at “XVIII edizione degli Incontri di Fisica delle Alte
    Energie”, Milano-Bicocca (Italy), 4-6 April 2018.} 
}
\instlist{
  \inst
  INFN, Sezione di Milano Bicocca - Milan, Italy.
  \inst
  Universit\`a degli studi di Milano-Bicocca - Milan, Italy.}

\begin{document}

\maketitle

\begin{abstract}
  The top quark is the heaviest elementary particle in the Standard Model (SM).
  For this reason, a precise determination of its mass $m_t$ is part of
  the LHC physics program. The most accurate determinations of $m_t$ rely on
  the kinematic reconstruction of the top decay products and on the use of
  Monte Carlo event generators.
  In these proceedings I discuss the impact of different
  theoretical descriptions for the process of top pair production at the LHC
  on the extraction of $m_t$.
\end{abstract}

\section{Introduction}
The phenomenology of the top-quark is driven by the large value
of its mass $m_t$. It is the only quark that decays instead of hadronizing,
giving us the opportunity to study the properties of a bare quark.
Due to radiative corrections, the value of $m_t$ has an impact on many
parameters of the SM like the Higgs self-couplings and  the bosons
masses.

The most accurate determinations of $m_t$ are the so-called ``direct
measurements''.
Monte Carlo generators are used to simulate templates of kinematic
distributions sensitive to the top quark mass. 
These templates are produced varying the input mass $m_t$ of the generator,
in order to extract the parametric dependence of the distributions on $m_t$.
The $m_t$ value that fits the data the best is the extracted top
quark mass.
This method has, however, an important drawback: the extracted $m_t$ value
strongly depends on the accuracy of the Monte Carlo~(MC) 
generator employed. This is a strong motivation for the theoretical community
to improve the accuracy of the MC generators.

In this contribution I compare the predictions of two
next-to-leading-order~(NLO) event generators for $t\bar{t}$
production implemented in the POWHEG framework
\cite{Nason:2004rx,Frixione:2007vw,Alioli:2010xd,Jezo:2015aia},
namely the \emph{hvq}~\cite{Frixione:2007nw} and the $b\bar{b}4\ell$
generators~\cite{Jezo:2016ujg} for some observables that can be used to
infer the top quark mass. These NLO generators must be matched with a
general purpose Shower Monte Carlo~(SMC) program, like {\tt
  Pythia8}~\cite{Sjostrand:2014zea}   or {\tt
  Herwig7}~\cite{Bahr:2008pv,Bellm:2015jjp}. I also show differences between 
distributions obtained with the two SMCs. The results presented here have
first been published in \cite{Ravasio:2018lzi}.

\section{Monte Carlo generators and simulated samples}

NLO (QCD) events for $t\bar{t}$ production in $pp$ collisions at a
center-of-mass energy $\sqrt{s}=8$~GeV have been produced using:
\begin{itemize}
  \item the \emph{hvq}~\cite{Frixione:2007nw} generator, which describes the
    process of $t\bar{t}$ production at NLO. The top decay is included at LO,
    using a reweighting procedure that allows to partially take into
    account off-shell effects and the spin correlation among $t$ and $\bar{t}$ decay
    products;
   \item the  $b\bar{b}4\ell$ generator~\cite{Jezo:2016ujg}, that implements
     the process $pp\to b\bar{b}e^+\nu_e\mu^-\bar{\nu}_\mu$ at NLO. Thus, as
     well as for top production, it describes its decay at NLO including
     off-shell and spin-correlations effects and the interference between radiation
     in production and in decay. Furthermore, all the processes which yield
     the same final state and their quantum interference are correctly included.
\end{itemize}
Selection cuts to suppress the $Wt$ topology, which is present only in
$b\bar{b}4\ell$, are adopted~\cite{Ravasio:2018lzi}.

In order to implement the $b\bar{b}4\ell$ generator, a resonance-aware
formalism has been developed, which is coded in the new {\tt
  POWHEG\,BOX\,RES} framework~\cite{Jezo:2016ujg}.
  By default, it generates multiple emissions, one
from the production process and one from each decayed resonance.
{\tt Pythia8}~\cite{Sjostrand:2014zea} (or {\tt
  Herwig7}~\cite{Bahr:2008pv,Bellm:2015jjp}) completes the NLO events, by
adding subsequent emissions in the soft and collinear approximation (provided by the parton
shower, PS) and translating the partonic final state into a hadronic one.
In order not to spoil the NLO accuracy of the result, the PS must veto
emissions that have a transverse momentum larger than the POWHEG radiation.
This is done per default both by {\tt Pythia8} and {\tt
  Herwig7} if the emissions come from the production process. We extended
this veto procedure also to the case of radiation from decayed top quarks.
Since {\tt Pythia8} implements a transverse momentum ordered
PS, it is enough to require that the first emission has a transverse
momentum smaller than the POWHEG emission.
{\tt Herwig7} is an angular ordered parton shower, thus all the emissions
must be checked.

\section{Comparison between the NLO generators showered with {\tt Pythia8}}

We choose simple observables $O$ that can be directly related to the top quark
mass:
\begin{equation}
  O =  O_c + B(m_t-m_t^c), 
\end{equation}
where $m_t^c$ is our reference mass $m_t^c=172.5$~GeV and $O_c$ and $B$ are
parameters that can be fitted using a MC generator. Neglecting the
differences among the $B$ coefficients between the generators, the
differences in the extracted mass can be written as
\begin{equation}
  \Delta m_t = - \frac{\Delta O_c}{B}.
\end{equation}

The first observable we examine is the peak of the invariant mass of the reconstructed
top quark ($W$ and $b$ jet system), \emph{i.e.} $m^{\max}_{Wb_j}$. In this case $\Delta m_t \approx -\Delta m^{\max}_{Wb_j}$.
Even if we apply a Gaussian smearing of 15~GeV to mimic
experimental resolution effects, we find a remarkable agreement between the peak
position extracted using $b\bar{b}4\ell$ and \emph{hvq}. 


It is also intriguing to have a look to purely leptonic observables, as
proposed by the authors of Ref.~\cite{Frixione:2014ala}. There is an overall
agreement between $b\bar{b}4\ell$ and \emph{hvq} predictions, apart from
observables sensitive to spin correlations, that are only approximatively
described at LO by \emph{hvq}. Two examples are given by
$\langle m(e^+\mu^-) \rangle$, which yields to $\Delta m_t \approx 1.5$~GeV,
and $\langle p_\perp(e^+\mu^-) \rangle$, which leads to
$\Delta m_t \approx-2$~GeV.

\section{Comparison between {\tt Herwig7} and {\tt Pythia8} predictions}

Quite large differences arise when we shower the events with {\tt Herwig7} 
instead of {\tt Pythia8}.
If we look at the reconstructed-top mass peak, {\tt Herwig7} prediction is
1~GeV smaller if $b\bar{b}4\ell$ is employed, 0.5~GeV in the case of
\emph{hvq}.
A worse agreement is found when leptonic observables are considered: the
$m_t$ value extracted with {\tt Herwig7} is roughly 2.5~GeV larger than the
{\tt Pythia8} one for both NLO generators.

This is not completely unexpected, given the different nature of the {\tt
  Herwig7} PS. 

\section{Conclusions}
When using {\tt Pythia8}, the differences between \emph{hvq} and
$b\bar{b}4\ell$ are large enough to use the newest generator, but not large
enough to completely overturn the current measurements that are based on
\emph{hvq}. 
When {\tt Herwig7} is used, we do not have a nice and consistent picture,
however we do believe that the option of dismissing {\tt Herwig7} 
is not soundly motivated. The difference between the two PSs may be due to
higher-order effects, that must be taken into account when estimating
theoretical uncertainties. When a realistic analysis is
performed, the parameter of a MC are tuned to reproduce the data fairly.
This could improve the agreement between {\tt Herwig7} and {\tt
  Pythia8}.

\acknowledgments
The author acknowledges T.~Je\v{z}o, P.~Nason and M.~Rocco for suggestions on the manuscript.


\begin{thebibliography}{0}


\bibitem{Nason:2004rx}
  P.~Nason,
  JHEP {\bf 0411} (2004) 040


\bibitem{Frixione:2007vw}
  S.~Frixione, P.~Nason and C.~Oleari,
  JHEP {\bf 0711} (2007) 070


\bibitem{Alioli:2010xd}
  S.~Alioli, P.~Nason, C.~Oleari and E.~Re,
  JHEP {\bf 1006} (2010) 043

 
\bibitem{Jezo:2015aia}
  T.~Je\v{z}o and P.~Nason,
  JHEP {\bf 1512} (2015) 065

\bibitem{Frixione:2007nw}
  S.~Frixione, P.~Nason and G.~Ridolfi,
  JHEP {\bf 0709} (2007) 126

\bibitem{Jezo:2016ujg}
  T.~Je\v{z}o, J.~M.~Lindert, P.~Nason, C.~Oleari and S.~Pozzorini,
  Eur.\ Phys.\ J.\ C {\bf 76} (2016) no.12,  691

\bibitem{Sjostrand:2014zea} 
  T.~Sj\"ostrand {\it et al.},
  Comput.\ Phys.\ Commun.\  {\bf 191}, 159 (2015)
  
\bibitem{Bahr:2008pv}
  M.~Bahr {\it et al.},
  Eur.\ Phys.\ J.\ C {\bf 58} (2008) 639
  
\bibitem{Bellm:2015jjp}
  J.~Bellm {\it et al.},
  Eur.\ Phys.\ J.\ C {\bf 76} (2016) no.4,  196

\bibitem{Ravasio:2018lzi}
  S.~Ferrario Ravasio, T.~Je\v{z}o, P.~Nason and C.~Oleari,
  Eur.\ Phys.\ J.\ C {\bf 78} (2018) no.6,  458


\bibitem{Frixione:2014ala}
  S.~Frixione and A.~Mitov,
  JHEP {\bf 1409} (2014) 012
  
  
\end{thebibliography}
\end{document}